# Angular Resolution Enhancement of Electron Backscatter Diffraction Patterns

Ben Britton* and Tianbi Zhang

Department of Materials Engineering, University of British Columbia, Vancouver, British Columbia, Canada

*ben.britton@ubc.ca

## Abstract

We present a simple 'shift and add' based improvement in the angular resolution of single electron backscatter diffraction (EBSD) patterns. Sub-pixel image registration is used to measure the (sub-pixel) difference in projection parameters for patterns collected within a map, and then the pattern is shifted and added together. The resultant EBSD-pattern is shown to contain more angular information than a long-exposure single pattern, via 2D Fast Fourier Transform (FFT)-based analysis. In particular, this method has the potential to enhance the scope of small compact direct electron detectors (DEDs).

## Introduction

Electron backscatter diffraction (EBSD) is a popular microanalysis technique in the scanning electron microscope [1], where the interaction between the primary electron beam and the specimen forms Kikuchi patterns. Each Kikuchi pattern contains diffraction information via a direct space projection of the crystal lattice, and features on the patterns such as Kikuchi bands and higher order Laue Zone (HOLZ) rings can be used for symmetry determination [2], phase classification [3,4] and orientation determination [5–10] through various pattern processing and analysis methods. Compared to spot diffraction patterns [11], EBSD patterns tend to have a higher proportion of pixels with crystallographic information encoded [12]. This means that good quality EBSD patterns often have a high density of subtle crystallographic information across high scattering angles. The quality of the signal in terms of both the signal to noise for each pixel but also the angular encoding of data across the pattern and therefore between pixels, ultimately determines the ultimate potential angular resolution and accuracy of the analysis method.

For EBSD pattern capture, the signal is generated after the electrons have scattered within the sample and struck the pixelated detector. Degradation of the signal can occur due to low signal to noise, e.g. due to high dark currents, inefficient collection or scattering within the detector. This can be measured by the modulation transfer function (MTF) and detection quantum efficiency (DQE) and direct electron detectors (DEDs) have been shown to typically have a higher DQE and MTF than indirect detectors [13,14]. Here we refer to a 'direct' electron detector as one that does not convert the signal to another form prior to measuring it (e.g. in a conventional indirect detector, diffracted electrons strike a scintillator which creates light and this light is later converted into digital signal, via a CCD or CMOS camera). A number of previous studies [15–23] have demonstrated the use of DEDs, including the use of hybrid pixel detectors (HPD) which are one of the predominant and more widely available DED technologies, to obtain high quality EBSD patterns using electron counting and energy thresholding.

Large and many-pixel DEDs are available, and are being used, in some laboratories for advancing EBSD analysis [21]. Additionally, there are more compact, and typically less expensive, detectors are also being employed and commercialized [24]. Many of these smaller and compact HPDs based upon the Medipix and Timepix families [25–27] of detector that have been initially developed by CERN and are now being used for electron detection. In brief, these HPD systems consist of a single crystal of doped silicon which is then exposed to the flux of incoming diffracting electrons. For the Timepix systems, on the back of the HPD are multiple (256 x 256 pixel) bump-bonded read outs which collect the charge from a local region of the single crystal detecting layer, and these are connected to an application specific integrated circuit (ASIC) to perform electron counting and/or integration of the incoming signal. For EBSD users, ultimate these imaging modes can be used to form an energy-filtered high quality EBSD pattern with a physical pixel size of 55 μm in a Timepix based HPD.

Despite the superior detector performance in theory, a particular concern of HPD devices is their pixel pitch (>25 μm) and typically smaller pixel array as compared both to (modern) indirect detector systems and alternative DEDs (e.g. MAPS-based devices [21]). These two factors can pose a limit on the ultimate angular resolution of the EBSD patterns captured and angular information encoded in them when compared to detectors with similar MTF and DQE performance but finer pixels.

In practice, when pixel pitch and pixel array size are fixed for a detector, placing the detector farther from the sample increases the angular detail captured by each pixel [28]. However, this comes at a cost of reduced total solid angle captured and may affect indexing and advanced analysis (such as full strain or rotation tensor evaluation [29]).

Further increase of pattern resolution should then resort to processing of patterns. It is noticed from previous high angular resolution (HR-) EBSD studies that information encoded on the pattern can be recovered at 1/10$^{th}$ to 1/20$^{th}$ of a pixel [30,31]. We note that the realized precision depends on detector type and acquisition parameters [32,33]. The use of large region-of-interest based analysis with fast Fourier transform based image analysis algorithms implies that it may be possible to increase the sampling of each individual pattern through the use of up-sampling, super-resolution or sub-pixel registration methods, by creating a super-resolution pattern using patterns with high quality yet lower number of pixels. Additionally, recent work using full pattern matching algorithms has also highlighted the potential of using a lower pixel number EBSD pattern to recover higher quality angular information [34].

To realize a single pattern with improved resolution, we can employ sub-pixel scheme and draw inspiration from the optics community where in 1980 Bates and Cady [35] used a 'shift-and-add' algorithm to improve the resolution of an image beyond the pixel pitch of the sensor. In this algorithm, multiple images are taken with small, and sub-pixel, spatial offsets of the camera and these are subsequently 'added together' via a summation scheme to achieve an increase in the resolution of the image.

In the present work, we exploit the fact that DED-based detector can collect a singular high-quality pattern due to the enhanced MTF of the detector, combined with energy filtering. Collection of multiple of these patterns can be used to enhance the quality inside the diffraction pattern further, via the application of a sub-pixel shift-and-add pattern processing to create a super-resolution EBSD pattern. We will demonstrate that the super-resolution pattern has further improved signal-to-noise ratio than a long-exposure, single frame EBSD pattern as well as patterns with high frame integration.

## Materials & Method

EBSD was performed in a Thermo Fisher Scientific Apreo 2 ChemiSEM with a TruePix EBSD detector at 20 keV landing energy. The TruePix detector is based on the Timepix chip with a Si sensor, which has 256x256 pixels with a pixel pitch of 55 µm. The detector operated in electron counting mode with an energy threshold of 19.4 keV (97% of the primary electron beam energy).

EBSD patterns were captured from a single crystal of semiconductor grade silicon mounted on an Al-stub, with the [001] direction out of plane and two corresponding {110} fracture surfaces aligned close to the vertical and horizontal beam scanning directions. The sample was mounted on the 45° pre-tilt holder of the multi purpose holder, and the stage was tilted a further 25° to tilt the sample at 70° in total.

EBSD data was collected with the sample mounted with a 11 mm working distance and a probe current of 3.2 nA.

Prior to pattern collection, flat fielding was performed by collecting multiple patterns at low magnification until a diffuse background was observed. Each recorded pattern was subsequently normalized (division) by this background during pattern acquisition.

Four experiments were performed:

(1) Standard detector distance – a 50 x 60 point map was collected with a 3 µm step size (i.e. 1/20 of a detector pixel size). Patterns were collected at an exposure time of 0.3 s. These patterns were used to create super-resolution patterns.
(2) Standard detector distance – increased pattern exposure, single frame acquisition. Variable exposure times were used. These patterns are used to evaluate the improvement of pattern quality by longer exposure time.
(3) Standard detector distance – a 15 x 15 point map was collected with a 50 nm step size (i.e. $1\times10^{-3}$ of a detector pixel size). Patterns were collected with an exposure time of 0.3 s. Summation of patterns from this map was used to mimic frame integration, which is used for pattern quality comparison.

(4) Camera retraction – the camera was retracted by 11.6 mm, to achieve a zoom of ~3x in the diffraction pattern, and a 5 x 5 point map was collected with a 3 μm step size. Each pattern was collected with an exposure time of 1.4 s. Summation of patterns from this map was used as a reference to compare with the super-resolution pattern.

Pattern quality is evaluated using a signal-to-noise ratio metric calculated from the radial profiles of the 2D fast Fourier transformation (FFT) of the EBSD patterns. The signal to noise was measured as the mean of the signal divided by the noise, where the signal was considered as information contained in the frequency range of between 2/128 and 96/128 cycles/pixel and noise was between 96/128 and 128/128 cycles/pixel.

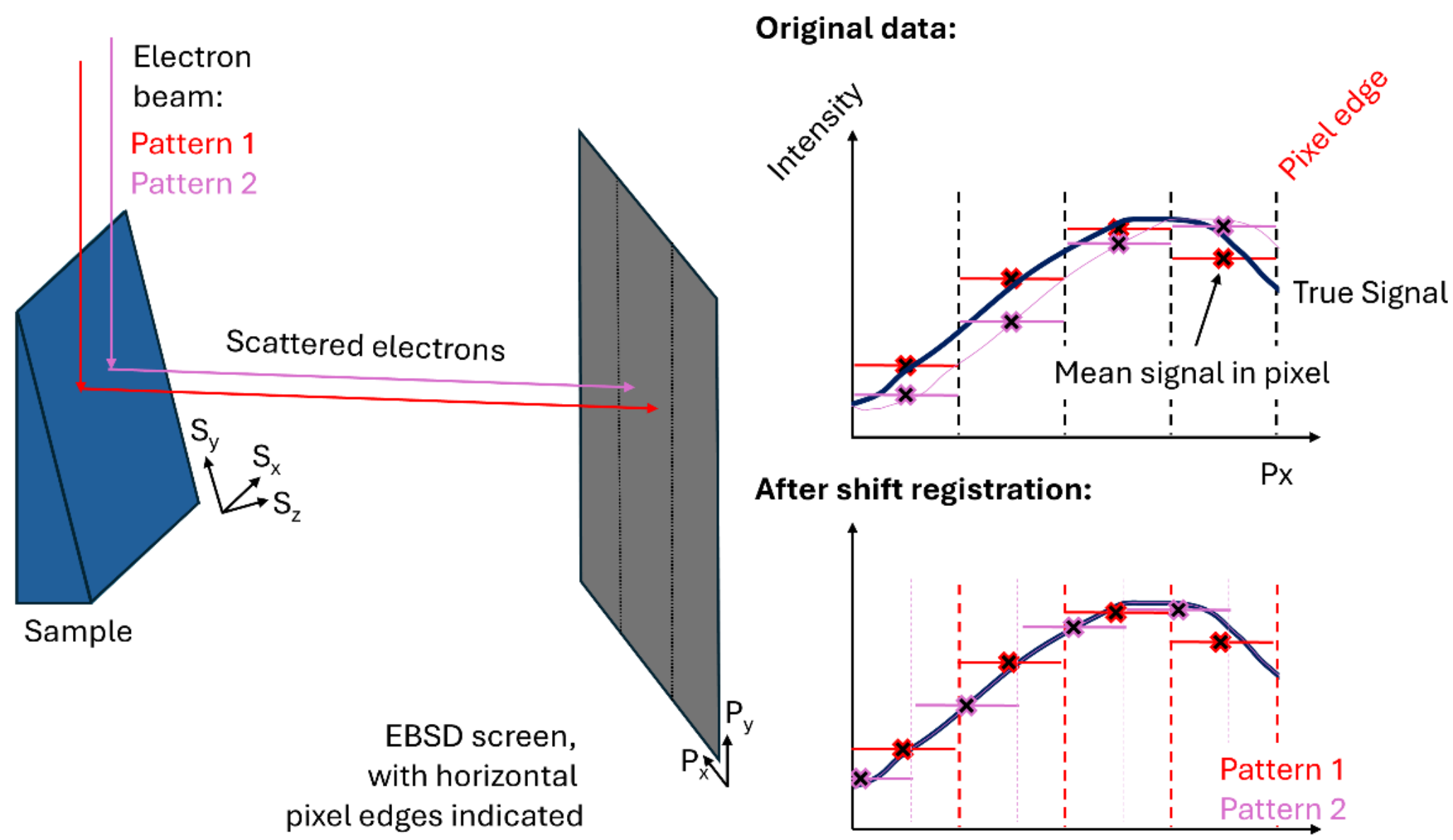

*Figure 1: Illustration of how a higher angular information can be reconstructed using shift registration of two patterns with a pixel displacement between the two pattern signals.*

The basic shift-and-add algorithm is illustrated in Figure 1. Two patterns are collected with a sub (detector) pixel offset between them, and image registration is performed prior to merging of the pattern data.

For map (1), shift-and-add-based sub-pixel resolution analysis was performed within MATLAB. This consisted of: (a) selection of a reference point in the map in the top centre; (b) measurement of the image transformation required to best align each pattern against this reference pattern, allowing for scale, rotation and translation degrees of freedom using phase-based image correlation with between every individual point in the '*imregcorr*' function. (c) subsequent transformation of each pattern into the frame of the reference pattern using the '*imwarp*' function, with bicubic interpolation; (d) summation of this image series as described later. The data and full code are provided for an interested reader to use (see the data availability statement), and the essential lines for this approach include:

```
[tform,t_peak] =
imregcorr(allProcessedPatterns(:,:,n_pat),SR_pattern_reference,Method="phasecorr");
```

```
temp_image=
imwarp(allProcessedPatterns(:,:,n_pat),tform,'bicubic',"OutputView",OutPutSize,FillValues=0);
```

Where *tform* is a rigid transformation function, that consists of translation, rotation, and scale terms; *t_peak* is the correlation peak height; *allProcessedPatterns* is the array containing the (background corrected) direct electron detector patterns in an array which is 256 x 256 x *n_pat* in size, such that *n_pat* is a counter for the pattern number; *SR_pattern_reference* is the super resolution reference pattern (i.e. the pattern that all the data is correlated against); *temp_image* is the output registered image, where the *allProcessedPatterns(:,:,n_pat)* is shifted back onto the reference pattern using the *tform* transformation. The image correlation is performed in the phase domain, and the warping is performed using a bicubic interpolation. The super resolution pattern can be created by summing multiple of '*temp_image*' arrays together depending on the exposure time. More details of the function syntax can be found within the MATLAB help files [36,37].

Processed patterns were saved to disc and analysed using scripts written in MATLAB and working with AstroEBSD [5]. Image processing was performed, as described shortly, in MATLAB. Radon-based pattern indexing [8] and refined template matching were used to index and align the patterns with high precision. Initial pattern matching was performed using a high-quality reference 'spherical pattern' as generated using eSprit DynamicS (Bruker Nano GmbH), with reprojection in AstroEBSD. A higher quality 'BWKD' simulation was created using the algorithms based within MapSweeper (Oxford Instruments) and a custom python front end. The higher quality pattern was simulated using 20 keV and a minimum $d_{hkl}$ of >0.25 angstrom and a $F_{hkl}$ of >5% for the strong beam portion of the simulation, to ensure that the higher order Laue zone rings and finer features would be reasonably reproduced.

## Results

Measurement of the image warping function is shown in Figure 2, with Map 1 data. The Peak Height map shows the quality of cross correlation for each warping function, and every point within the map has a high correlation score (>0.3). The autocorrelation point has the highest peak height (0.56). The peak height map reveals a square motif with each 'block' being ~55 µm in X-width, i.e. the physical size of the detector pixels. The Y-width is slightly larger, and this is due to the sample tilt as a shift down the sample, $S_y$ in Figure 1, results in a smaller shift down in the detector, $P_y$ in Figure 1, due to the sample tilt. The measured warping function shows very little image rotation (equivalent to <0.04°). As the sample Y position increases, the scale and Y translation values increase systematically (as expected), and as the X Position increases only the X translation values increase.

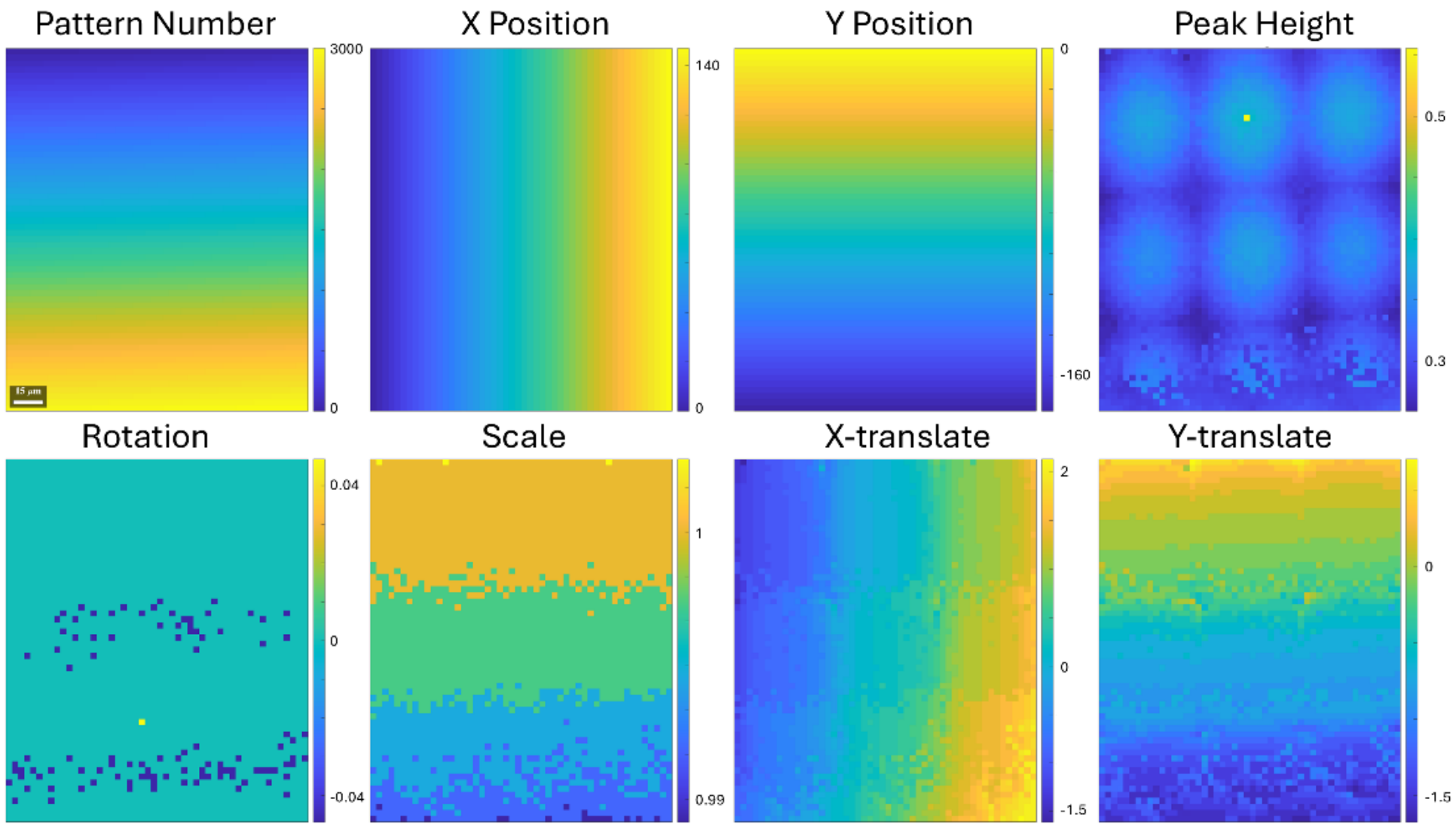

*Figure 2: Results from the image-based cross correlation of the diffraction patterns in the map. The reference point can be see as the 'bright' cross correlation point in the Peak Height map.*

Evaluation of the signal to noise with increasing pattern exposure time (i.e. dose), and the impact of the shift-and-add based super-resolution method are presented in Figure 3. A high quality pattern is obtained for the shortest recorded exposure time (18 ms), and an increase in exposure time increases the quality of the obtained pattern. This improvement can be observed both directly within the pattern, but also within high frequency components within the $\log_{10}$ 2D Fourier Power Spectrum based analysis. Quantitative analysis of the signal-to-noise ratio reveals that there is an immediate increase in signal to noise for increasing exposure time, and single patterns collected via variable exposures overlap well with patterns obtained from frame integration (using Map 2). However, the improvement in signal to noise plateaus at longer exposures towards a limit of a signal to noise of 30.

The shift-and-add-based super-resolution method increases quality of the patterns, as there is more information within the higher frequency domain (i.e. the Fourier transform 'spikes' in the 2D log polar transform extend further towards the edge of the window), and this is also reflected in the associated signal to noise ratio plot. Here multiple (effective) exposure times are taken from a random sub-set of the total registered Map 1 data set. Saturation of the quality of the super-resolution data occurs with an exposure time of ~1s.

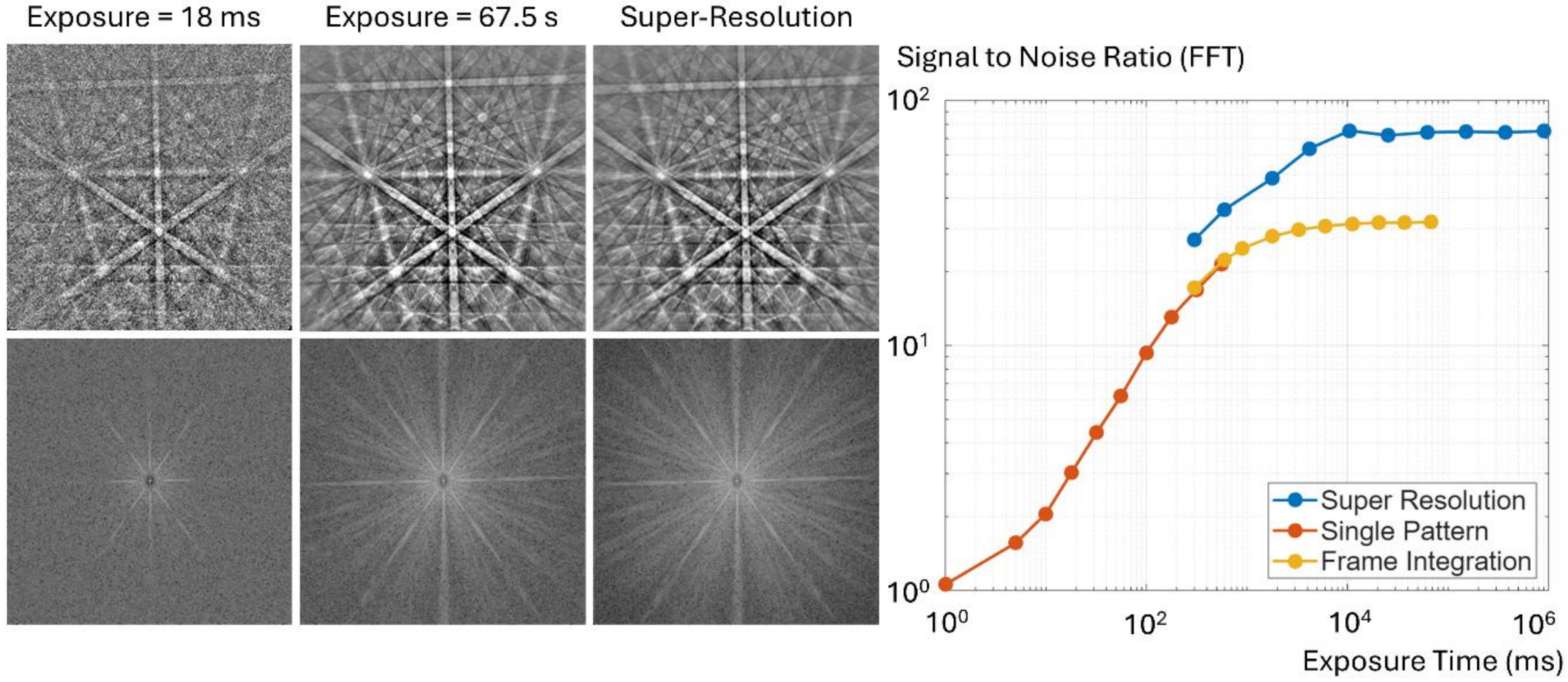

*Figure 3: Signal to noise analysis for EBSD patterns collected using single pattern acquisition with increasing exposure times; frame integration, and application of the super-resolution method.*

Next, the image warping functions were used to create an up-sampled diffraction pattern that matches the camera retract data (Map 4), and the comparison is shown in Figure 4. Additionally, this pattern is displayed alongside a high quality BWKD simulation and a simple *imresize* of a 1 frame pattern (performed with cubic interpolation). The super resolution pattern clearly includes many comparable features to both the camera retract and BWKD simulation, for example: (a) the HOLZ rings around the central <110> pole (found at the intersection of the {220} and {111} bands}; (b) the two <123> poles and their HOLZ rings to the left and right (found at the intersection of the {422} and {331} bands); (c) the fine structure within the vertical {220} band, as well as the higher order {440} and {660} edges.

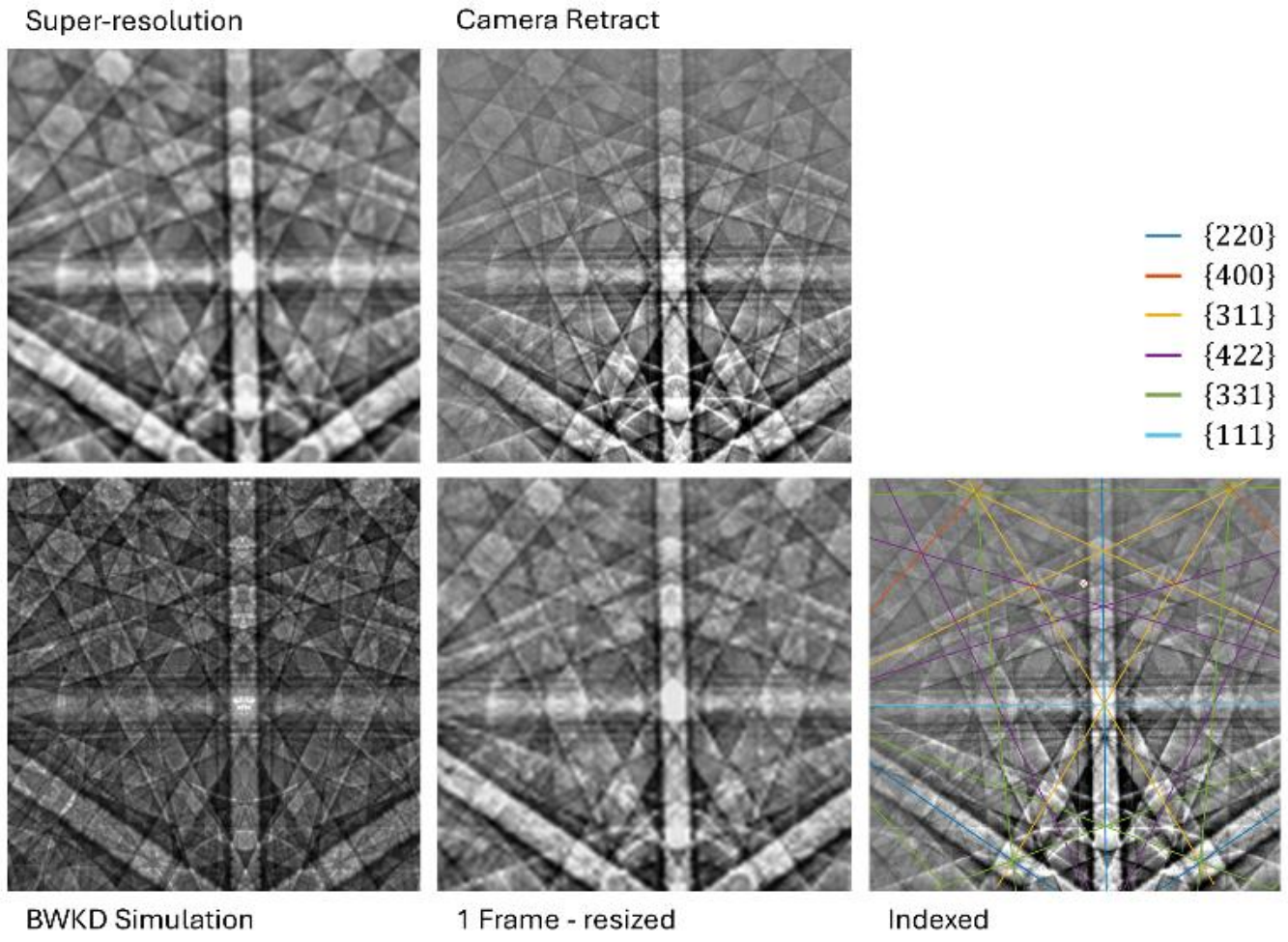

*Figure 4: Comparison of the super resolution pattern with camera retraction, BWKD simulation, and a single frame resized using a bicubic interpolation scheme.*

## Discussion

High quality EBSD patterns have been obtained using a combination of a high energy threshold, together with a shift-and-add algorithm to improve the angular information contained within the individual pattern. This increases the quality of the pattern as measured via a Fourier based signal to noise evaluation which reveals an increase in a signal-to-noise ratio of 30 to 70 (as per Figure 2) for an effective exposure time of 1 second per pattern.

The impact of this improvement can be observed within Figure 5, where the up-sampled Super Resolution pattern is now represented via application of the imaging warp functions to provide access to a pattern within a 2560 x 2560 pixel grid (i.e. a 10x up-sampling, from the original 256 x 256 pixel image). This pattern shows significant detail and also provides very high pattern quality at the corners of the pattern. It is worth noting that improved pattern quality at the corners of the pattern is especially important for indexing lower symmetry materials and addressing pseudosymmetry concern, as patterns that contain multiple high quality and well dispersed zone axes are always indexed more successfully with EBSD.

At first glance the 2D FFT-based power spectrum analysis of this up-sampled super resolution pattern in Figure 5 may look strange with the large vertical and horizontal band extending from the central region and no information within the corners. However this is expected from the Fourier-based analysis as it is an artifact of the up-sampling routine because no further information can be extracted beyond the original frequency domain, without the use of additional assumptions. The equivalence of the information can be confirmed as the insert zoom in of this central (256 x 256 pixel) region shows an exact copy of the original super resolution pattern. In practice, up-sampling like this is not required if pattern analysis is carried out within the Fourier domain, but it may be of use for pattern analysis in the real domain.

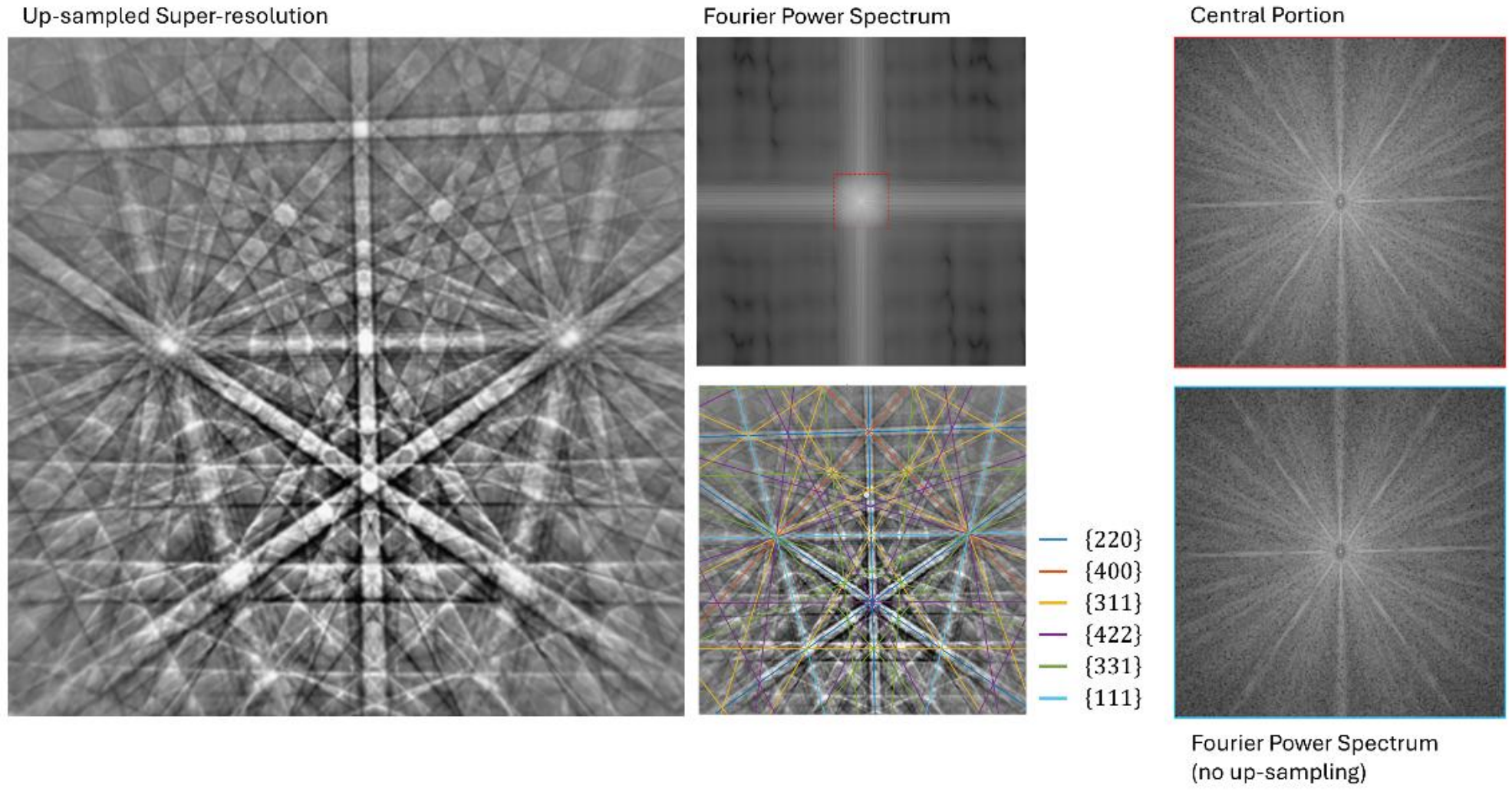

*Figure 5: Application of a ten-times pixel upscaling of the super resolution pattern, with pattern indexing and analysis of the Fourier power spectrum. The original pattern with no-upsampling, insert and indexed, was collected as a 256x256 frame.*

This manuscript has shown an up-sampling approach for the ‘simple’ Si-single crystal, and it relies on a common orientation and lattice within all the points in the map used for up-sampling. In practice there may be some methods to collect data from real samples which are not as simple as this:

(1) A large grain sample (c. 1 detector pixel in grain diameter) where the lattice is identical across the grain.
(2) Sub-detector pixel magnitude stage movements, together with collection from the sample physical point.
(3) Additional modification to the camera to provide sub-pixel movements of the detector image plane.

Further enhancement of the diffraction pattern could also be obtained if there are assumptions made about the nature of the blurring, similar to efforts within the ptychography community, to enhance the spatial resolution of 4D-STEM measurements through iterative refinement of the shape of the probe. In effect this could be achieved by estimating the point spread function (PSF) of the electron probe (and detector) [38]. However, note that that EBSD patterns collected here have been energy filtered to be close to the landing energy of the primary electron beam, which may impact the estimated shape of the interaction volume for the collection of this signal.

As alternative strategies to those presented here, high (angular) resolution methods could be used to measure the warping functions (e.g. applying the HR-EBSD algorithms more directly to measure the pattern shift and zoom terms, and/or variations in the crystal beyond simple pattern centre movements). Development of these algorithms may need to be careful with these approaches if there is a change in the volume of the unit cell (as this will blur the band edge locations) or significant movements of the patterns across the screen which can result in pattern blurring and variation in illumination of specific features due to physical effects, such as the excess deficiency modulation of the upper and lower edge of all the bands [12].

In this manuscript, the work has focussed on up-sampling very high-quality DED patterns as they have a high DQE at high spatial frequency to begin with, the shift-and-add algorithm can also be used for indirect patterns. Care should be taken in evaluating the performance of the up-sampling approach, as it may be that the pixel array of the detector is already sampling at a higher resolution than the diffraction pattern contains information [39].

The present manuscript demonstrates that the simple shift-and-add algorithm improves the angular information within the reconstructed EBSD pattern. Applications of this method could be extended towards the selection of an optimum 'reference pattern' for pattern comparison methods, including HR-EBSD strain and lattice rotation measurements, as well as evaluation of finer features within the pattern that may be important for unit cell determination, centrosymmetric evaluation, and addressing indexing challenges associated with pseudosymmetry.

## Conclusion

The signal to noise of a 256x256 pixel direct electron detector based EBSD pattern has been increased by more than a factor of two, via the application of a sub-pixel shift-and-add pattern reconstruction algorithm.

This higher quality super resolution pattern has been compared with single patterns captured with ever increasing intensities, and it is superior to a pattern collected with an equivalent dose.

This simple method can be applied with no hardware modification and can be used to improve pattern analysis further.

## Acknowledgements

We would like to thank Dr. Aimo Winkelmann (AGH University of Kraków) for kindly providing the BWKD simulation package. We acknowledge the following funding support: Natural Sciences and Engineering Research Council of Canada (NSERC) [Discovery grant: RGPIN-2022-04762, 'Advances in Data Driven Quantitative Materials Characterization'], Canada Foundation for Innovation and British Columbia Knowledge Fund (BCKDF) via the John R. Evans Leaders Fund (CFI-JELF) [#43737, 3D-MARVIN]. We thank Graeme Francolini for providing helpful comments on a draft of the manuscript.

## Credit Author Contribution Statement

T. Ben Britton: conceptualization, formal analysis, methodology, funding acquisition, project administration, resources, supervision, visualization, validation, writing – initial draft

T. Zhang: data curation, formal analysis. writing – review & editing.

## Data Availability Statement

EBSD data are available on Zenodo at 10.5281/zenodo.17905368. The MATLAB code 'deck' for analysis can be found on github:
https://github.com/ExpMicroMech/AstroEBSD/main/decks/SinglePatchEBSP_Plus_SNR.m.